\documentclass[prd,twocolumn,tightenlines,superscriptaddress,nofootinbib,
showpacs]{revtex4}
\usepackage{amssymb,latexsym}
\usepackage{amsmath,amsbsy,bbm}
\usepackage{epsfig,bm}
\usepackage{graphicx,comment}
\usepackage{color}
\usepackage{soul}
\usepackage[normalem]{ulem}
\begin{document}

\title{Universal scaling of N\'eel temperature, staggered magnetization density, 
and spinwave velocity of three-dimensional disordered and clean quantum 
antiferromagnets}

\author{D.-R. Tan}
\affiliation{Department of Physics, National Taiwan Normal University,
88, Sec.4, Ting-Chou Rd., Taipei 116, Taiwan}
\author{F.-J. Jiang}
\email[]{fjjiang@ntnu.edu.tw}
\affiliation{Department of Physics, National Taiwan Normal University,
88, Sec.4, Ting-Chou Rd., Taipei 116, Taiwan}

\begin{abstract}
The N\'eel temperature, staggered magnetization density, as well as the 
spinwave velocity of a three-dimensional (3D) quantum Heisenberg model with
antiferromagnetic disorder (randomness) are calculated using first principles 
non-perturbative quantum Monte Carlo simulations. In particular,
we examine the validity of universal scaling relations that are related to 
these three studied physical quantities. These relations are relevant to
experimental data and are firmly established 
for clean (regular) 3D dimerized spin-1/2 Heisenberg models. 
Remarkably, our numerical results show that the considered scaling relations 
remain true for the investigated model with the introduced disorder.
In addition, while the presence of disorder may change the 
physical properties of regular dimerized models, hence leading to
different critical theories, both the obtained data of 
N\'eel temperature and staggered magnetization density in our study are fully 
compatible with the expected critical behaviour 
for clean dimerized systems. As a result, it is persuasive to conclude that the related 
quantum phase transitions of the considered 
disordered model and its clean analogues are governed by the same critical 
theory, which is not always the case in general.  
Finally, we also find smooth scaling curves 
even emerging when both the data of the investigated disordered model as well 
as its associated clean system are taken into account concurrently. This in turn implies that, 
while in a restricted sense, the considered scaling relations for 3D spin-1/2 
antiferromagnets are indeed universal.
   
\end{abstract}


\maketitle

\section{Introduction}
Universality is an elegant concept and frequently appears in all fields of 
physics in various forms. In addition to being important in theoretical 
physics, the idea of 
universality can also serve as useful guidelines for experiments. 
One well-known example of the usefulness of universality
is the critical exponents of second order phase 
transitions \cite{Nig92,Sac99,Car10}.
Specifically, the numerical values of critical exponents,
such as $\nu$ related to the correlation length and $\beta$ associated with
the magnetization, do not in principle depend on the microscopic details
of the underlying models, but are closely connected to the symmetries of the
considered systems. For example, the zero temperature phase transitions 
of two-dimensional (2D) dimerized quantum Heisenberg models are governed
by the $O(3)$ universality class \cite{Mat02,Wan05,Wen08,Alb08,Wen09,Jia09,Jia12}, 
which is originally resulted from the three-dimensional (3D) classical 
Heisenberg model \cite{Cam02,Pel02}. Furthermore, the spatial dimerization patterns have no impact
on the critical theories of these quantum phase transitions (However,
there may be anomolous corrections to scalings \cite{Wen08,Fri11,Jia12}). Another 
noticeable kind (and example) of universality is the generally 
applicable finite-temperature and -volume expressions of several physical
quantities of antiferromagnets 
\cite{Cha88,Reg88,Cha89,Neu89,Has90,Has91,Has93,Chu94,Wie94,Tro95,Bea96,San97,Jia08,Jia09.1,Jia11.1}. 
To be more precise, based on the corresponding low-energy effective
field theory, the theoretical predictions of these observables,
such as staggered and uniform susceptibilities, depend 
solely on a few parameters and have the same forms regardless of the 
magnitude of the spin of the systems. In conclusion, universality does play 
a crucial role in major areas of physics.    
   
Recently, the experimental data of the phase diagram of 
TlCuCl$_3$ under pressure \cite{Cav01,Rue03,Rue08} 
have triggered many studies both theoretically and
experimentally. In particular, several universal scaling relations are 
established for 3D quantum 
antiferromagnets \cite{Kul11,Oit12,Jin12,Kao13,Mer14}. Specifically, 
near the quantum phase transitions
of clean (regular) 3D dimerized spin-1/2 Heisenberg models, the N\'eel temperatures $T_N$ 
scale in several universal manners with the corresponding staggered 
magnetization 
density $M_s$ regardless of the dimerization patterns. In addition,
a quantum Monte Carlo study conducted later demonstrates that these
universal scaling relations even remain valid when (certain kinds of) 
quenched bond disorder, i.e. antiferromagnetic bond randomness are introduced into the 
systems \cite{Tan15}. Notice
the upper critical spatial dimension of the mentioned zero-temperature
phase transitions is three. Consequently, close to the critical points 
one expects to observe multiplicative logarithmic corrections to 
$T_N$ and $M_s$ when 
these quantities are considered as functions of the strength of dimerization.
Very recently an analytic investigation even argues that the widely believed
phase diagram of 3D quantum antiferromagnets is modified dramatically due
to these logarithmic corrections \cite{Har16}.   
The exponents related to these logarithmic corrections are determined
analytically in Refs.~\cite{Ken04,Ken12,Har15,Yan15}. Furthermore, the theoretical
predictions of the numerical values of these new exponents have been 
verified as well \cite{Yan15}.
Notice that the exponents related to the logarithmic corrections
to $T_N$ and $M_s$ take the same values for three spatial dimenions.
Using this result as well as the fact that the $\beta$ and $\nu$
have the same mean-field values, it is straightforward to show
that close the the phase transition, as functions of their
corresponding $M_s$, $T_N/c^{3/2}$ and $T_N/\overline{J}$ are linear in $M_s$
without any logarithmic corrections. 
Here $c$ and $\overline{J}$ are the low-energy constant spinwave velocity 
and the average of antiferromagnetic couplings, respectively.
These connections between $T_N$ and the associated $M_s$, namely 
$T_N/c^{3/2} = AM_s$
and $T_N/\overline{J} = A_1M_s$ ($A$ and $A_1$ are some constants) in the 
vicinity of a quantum critical point are confirmed 
in Ref.~\cite{Yan15}. In real materials, impurities are often 
present \cite{Vaj02}.
In addition, studies of quenched disorder effects on Heisenberg-type models 
continue to be one of the
active research topics in condensed matter physics \cite{Voj10}. Therefore one 
intriguing physics to explore further is that whether
the logarithmic corrections, as well as the linear dependence of $T_N/c^{3/2}$ 
and $T_N/\overline{J}$ on their associated $M_s$ are valid
for 3D systems with the presence of antiferromagnetic bond 
disorder. Since such studies of disordered models are relevant to
the experimental data of TlCuCl$_3$, in this investigation we have carried out 
a large-scale quantum Monte Carlo simulations of a 3D spin-1/2 antiferromagnet
with configurational disorder which is first introduced in Ref.~\cite{Yao10}. 
Remarkably, our data indicate convincingly that for the studied disordered model, 
$\overline{T_N}/\overline{c}^{3/2}$ and $\overline{T_N}/\overline{J}$
do depend on their corresponding $\overline{M_s}$ linearly close to the associated 
quantum phase transition (In this study observables with a overline on them 
refer to the results of disorder average). Furthermore, the obtained data of 
$\overline{T_N}$ and $\overline{M_s}$ here can be 
described well by the expected critical behaviour for 
regular dimerized models. This suggests that
the related quantum phase transition of the considered disordered system may 
be governed by the same critical theory as that of its clean analogues. 
It should be pointed out that while close to the considered quantum critical
point the relations
$\overline{T_N}/\overline{c}^{3/2} = A \overline{M_s}$ and 
$\overline{T_N}/\overline{J} = A_1\overline{M_s}$ hold even for the 
investigated model with the employed disorder, based on the results of 
current and previous studies, we find that the prefactor $A$ and $A_1$
are likely to be model-dependent. Surprisingly, when both the data of current
study and that of a clean system available in Ref.~\cite{Kao13} are taken into
account concurrently, smooth universal curves appear. This in turn implies
that, while in a restricted sense, the investigated scaling relations of 
3D spin-1/2 antiferromagnets are indeed universal.
Finally, although the prefactor $A$ of $T_N/c^{3/2} = A M_s$ 
($\overline{T_N}/\overline{c}^{3/2} = A \overline{M_s}$) is likely 
not universal, we find the quantity $A/(J'/J)_c^{3/2}$, where
$(J'/J)_c$ is the considered critical point, obtained in this study matching
very well with the corresponding one in 
Ref.~\cite{Yan15}. This indicates that with an appropriate normalization, 
a true universal quantity may still exist. 

This paper is organized as follows. After the introduction,
in section 2 we define the investigated model as well as the calculated
observables. We then present a detail analysis of our numerical data
in section 3. In particular, the scaling relation $T_N/c^{3/2} = A M_s$
as well as the logarithmic corrections to $T_N$ and $M_s$ are examined
carefully. Finally in section 4 we conclude our study.  

\vskip-1.055cm

\section{Microscopic model, configurational disorder, and observables}
\begin{figure}
\begin{center}
\includegraphics[width=0.3\textwidth]{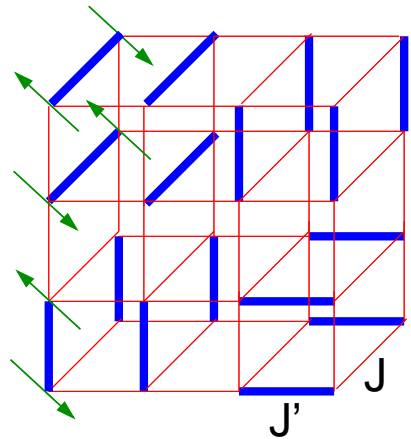}
\end{center}
\caption{The model with configurational disorder studied here.}
\label{model_fig1}
\end{figure}

The 3D quantum Heisenberg model with random antiferromagnetic couplings
studied here is given by the Hamilton operators
\begin{eqnarray}
\label{hamilton}
H = \sum_{\langle ij \rangle}J_{ij}\,\vec S_i \cdot \vec S_{j} 
+ \sum_{\langle i'j' \rangle}J'_{i'j'}\,\vec S_i \cdot \vec S_{j} ,
\end{eqnarray}
where $J_{ij}$ and $J'_{i'j'}$ are the antiferromagnetic couplings (bonds)
connecting nearest neighbor spins $\langle  ij \rangle$ and
$\langle  i'j' \rangle$, respectively, and $\vec{S}_i$ is
the spin-1/2 operator at site $i$. The bond disorder considered in this 
investigation is a generalization of the configurational disorder introduced
in Ref.~\cite{Yao10} and is realized here as follows.
First of all, a given cubical lattice is subdivided into 
two by two by two cubes. 
Secondly, the 12 bonds within a cube is classified into 
three sets of bonds so that each of them is made up of four bonds parallel to 
a particular coordinate axis.
Furthermore, one of the three sets of
bonds of every cube is chosen randomly and uniformly. 
In particular, these picked bonds are assigned the antiferromagnetic coupling 
strength $J'$. Finally the remaining unchosen bonds as well as
those not within any cubes have antiferromagnetic coupling strength $J$ 
which is set to be $1.0$ in this investigation. 
Figure~\ref{model_fig1} demonstrates one realization of the model with 
configurational disorder studied here. Notice in our study the couplings 
$J'$ and $J$ satisfy $J' > J$. Hence as the ratio of $J'/J$ increases, 
the system will undergoes a quantum phase transition. 

To determine the N\'eel temperature 
$\overline{T_N}$, the staggered magnetization density $\overline{M_s}$,
as well as the spinwave velocity $\overline{c}$ of 
the considered models with the employed configurational disorder, 
the observables staggered structure factor $\overline{S(\pi,\pi)}$, 
both the spatial and temporal winding numbers squared 
($\overline{\langle W_i^2 \rangle}$ for $i \in \{1,2,3\}$ and
$\overline{\langle W_t^2 \rangle}$), spin stiffness $\overline{\rho_s}$,
first Binder ratio $\overline{Q_1}$, and second Binder ratio $\overline{Q_2}$ 
are calculated in our simulations. 
The quantity $\overline{S(\pi,\pi)}$ takes the form 
\begin{equation}
\overline{S(\pi,\pi)} = 3 \overline{\langle ( m_s^z )^2\rangle}
\end{equation}
on a finite cubical lattice with linear size $L$. Here 
$m_s^z = \frac{1}{L^3}\sum_{i}(-1)^{i_1+i_2+i_3}S^z_i$ with $S^{z}_i$ being
the third-component of the spin-1/2 operator $\vec{S}_i$ at site $i$. 
In addition, the spin stiffness $\overline{\rho_s}$ has the following expression
\begin{equation}
\overline{\rho_s} = \frac{1}{3\beta L}\sum_{i=1,2,3}\overline{\langle W_i^2 \rangle},
\end{equation}
where $\beta$ is the inverse temperature.
Finally the observables $\overline{Q_1}$ and $\overline{Q_2}$ are defined by
\begin{equation}
\overline{Q_1} = \frac{\overline{\langle |m_s^z| \rangle }^2}{\overline{\langle (m_s^z)^2\rangle}} 
\end{equation}
and
\begin{equation}
\overline{Q_2} = \frac{\overline{\langle (m_s^z)^2 \rangle }^2}{\overline{\langle (m_s^z)^4\rangle}},
\end{equation}
respectively. With these observables, the physical quantities required for our 
study, namely $\overline{T_N}$, $\overline{M_s}$, and $\overline{c}$, can be 
calculated accurately.

\begin{figure}
\begin{center}
\includegraphics[width=0.38\textwidth]{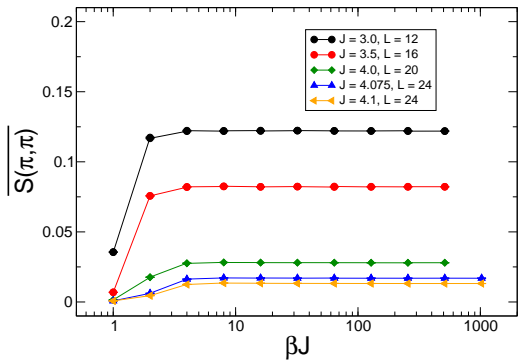}
\end{center}\vskip-0.5cm
\caption{Convergence of the structure factors $\overline{S(\pi,\pi)}$ to their
ground state values for several considered $J'/J$ and box sizes $L$. 
The solid lines are added to guide the eye.}
\label{ms_fig1}
\end{figure}

\section{The numerical results}
To examine whether the scaling relations 
$\overline{T_N}/\overline{c}^{3/2} = A\overline{M_s}$ and 
$\overline{T_N}/\overline{J} = A_1\overline{M_s}$, where $A$ 
and $A_1$ are some constant, 
appear for the considered 3D quantum Heisenberg models with the introduced 
configurational disorder, we have carried out a large-scale Monte Carlo 
simulation using 
the stochastic series expansion (SSE) algorithm with very efficient 
loop-operator update \cite{San99}.
We also use the $\beta$-doubling scheme \cite{San02} 
in our simulations so that $\overline{M_s}$ can be obtained efficiently. 
Here $\beta$ refers to the inverse temperature. Furthermore,
each disordered configuration is generated by its own random seed in order
to reduce the effect of correlation between observables determined from
different configurations. Our preliminary results indicate that
the critical point $(J'/J)_c$ lies between 4.15 and 4.17. Hence 
we have focused on the data of $J'/J \le 4.13$.
Notice in our study, $\overline{M_s}$ are calculated using several hundred 
configurations and 
$\overline{T_N}$ ($\overline{c}$) are determined with several thousand
(few to several ten thousand) disorder realizations.
The convergence of the considered observables to their correct values 
associated with the employed disorder, 
as well as the systematic uncertainties due to Monte Carlo sweeps within 
each randomness 
realization, number of configurations used for disorder
average, and thermalization are examined by performing many trial simulations
and analysis. The resulting data from these trial simulations and analysis
agree quantitatively with those presented here. Notice the statistics reached 
for studies of clean systems typically is better than that
of investigation related to disordered models. Therefore, we have additionally 
carried out many calculations using exactly the same parameters to 
estimate the uncertainties due to the statistics obtained here.
In summary, the quoted errors of the calculated observables in this study 
are estimated with conservation so that the influence of these mentioned 
potential systematic uncertainties are not underestimated.     

\subsection{The determination of $\overline{M_s}$}
\begin{figure}[ht!]
\begin{center}
\includegraphics[width=0.38\textwidth]{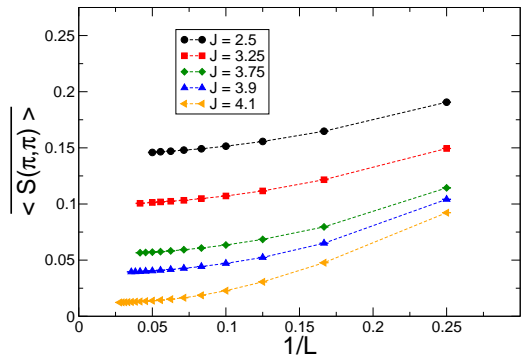}
\end{center}
\caption{$1/L$-dependence of the staggered structure factors 
$\overline{S(\pi,\pi)}$ for several considered values of $J'/J$.
The dashed lines are added to guide the eye.}
\label{ms_fig2}
\end{figure}

\begin{figure}
\begin{center}
\includegraphics[width=0.38\textwidth]{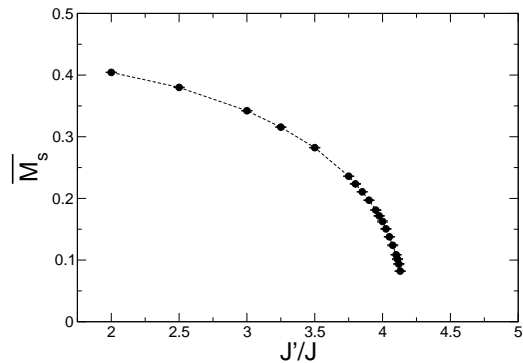}
\end{center}
\caption{$\overline{M_s}$ as functions of the considered values of
$J'/J$. The dashed lines are added to guide the eye.}
\label{ms_fig3}
\end{figure}

The observable considered here for the calculations of 
$\overline{M_s}$
is $\overline{S(\pi,\pi)}$. Specifically, for a given $J'/J$, the related
$\overline{M_s}$ is given by the square root of the corresponding bulk 
$\overline{S(\pi,\pi)}$. It should be pointed out that the zero temperature,
namely the ground state values of $\overline{S(\pi,\pi)}$ are needed for these 
calculations. Hence the $\beta$-doubling scheme is used here. The 
$\beta$-dependence, i.e. inverse temperature-dependence of 
$\overline{S(\pi,\pi)}$ for several considered $J'/J$ and $L$ is shown in 
fig.~\ref{ms_fig1}.
In addition, the $1/L$-dependence of the ground state $\overline{S(\pi,\pi)}$ 
for some studied $J'/J$
is depicted in fig.~\ref{ms_fig2}. The largest box size reached here for 
calculating the staggered structure factors is $L = 36$. 
Motivated by the theoretical predictions in Ref.~\cite{Car96}, 
the determination of $\overline{M_s}$ is done by 
extrapolating the related finite volume staggered structure factors 
to the corresponding bulk results, using the following four ansatzes
\begin{eqnarray}
\label{poly3}&&a_0 + a_1/L + a_2/L^2 + a_3/L^3, \\
\label{poly2}&&b_0 + b_1/L + b_2/L^2, \\
\label{poly31}&&c_0 + c_2/L^2 + c_3/L^3, \\
\label{poly21}&&d_0 + d_2/L^2.
\end{eqnarray}
In particular, $\overline{M_s}$ are obtained by taking the square root of
the resulting constant terms of the fits. 
Figure \ref{ms_fig3} shows the numerical values of $\overline{M_s}$ which
are determined from the fits employing ansatz (\ref{poly2}) and
are used for the required analysis in this investigation. Furthermore, we 
also make sure that the presented results in fig.~\ref{ms_fig3} are convergent 
with respect to the elimination of more smaller box size data in the fits, 
and are consistent with the stablized values obtained from applying 
ansatzes (\ref{poly3}), (\ref{poly31}), and (or) (\ref{poly21}) to fit 
the data.  

\vskip0.5cm

\begin{figure}
\vskip0.25cm
\begin{center}
\includegraphics[width=0.38\textwidth]{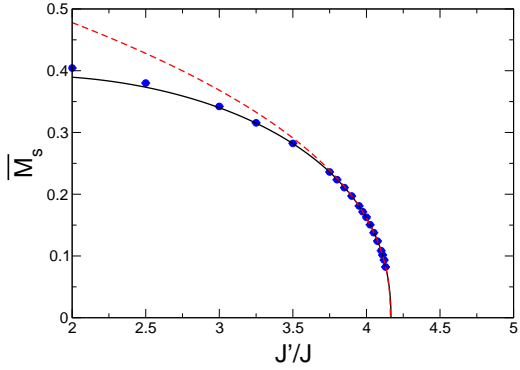}
\end{center}
\caption{Fits of $\overline{M_s}$ to the ansatzes of Eq.~(\ref{ms_log}) 
(solid line) and a pure power function $a_1|j_c-j|^{b_1}$ (dashed line). 
The range of $J'/J$ and $\chi^2/{\text{DOF}}$ for the fit using Eq.~(\ref{ms_log}) 
(ansatz $a_1 |j_c-j|^{b_1}$)
are $J'/J \ge 3.75$ and 1.2 ($J'/J \ge 3.9$ and 1.1), respectively.
The leading exponent $b$ of Eq.~(\ref{ms_log}) ($b_1$ of $a_1 |j_c-j|^{b_1}$)
determined from the fit is 0.513(6) (0.42(1)).
Applying the ansatz $a_1 |j_c-j|^{b_1}$ (Eq.~(\ref{ms_log})) to fit the data containing
those of $J'/J \le 3.5$ ($J'/J \le 2.5$) leads to very poor fitting quality. }
\label{ms_fig4}
\end{figure}

In Ref.~\cite{Yan15}, the 3D dimerized double cubic quantum 
Heisenberg model is studied. 
In particular, the relation of $T_N/c^{3/2} = A M_s$ is examined in 
detail.
Since three spatial dimensions is the upper critical dimension of the quantum 
phase transition considered in Ref.~\cite{Yan15}, one expects to observe 
logarithmic corrections to $M_s$ and $T_N$ when approaching the critical point. 
The theoretical calculations of the critical exponents associated with these
logarithmic corrections are available in Refs.~\cite{Ken04,Ken12,Yan15}, and 
the predicted values are
confirmed by a careful analysis of $M_s$ and $T_N$ conducted in Ref.~\cite{Yan15}.
Since disorder may change the upper critical dimension of the clean system,
it will be interesting to check whether this is indeed the case for our model.
The exponent related to the logarithmic correction to $M_s$, namely $\hat{\beta}$ 
has a value of $\frac{3}{11}$ for 3D clean dimerized model. Inspired by 
this, we have fitted our $\overline{M_s}$ data to an ansatz of the form 
\begin{equation}
a|j_c-j|^{b}|\ln(|(j_c-j)/j_c|)|^{3/11},
\label{ms_log}
\end{equation}
where $j = J'/J$ ($j_c = (J'/J)_c$ is the critical point) and $a$ is a 
constant. Notice the $b$ appearing in Eq.~(\ref{ms_log}) is the associated 
leading critical exponent which
is predicted to be 0.5. Interestingly, we find that the numerical values of 
$b$ (in Eq.~(\ref{ms_log})) obtained from the fits have an average
of 0.507(18) which is in reasonable good 
agreement with the predicted mean-field result $0.5$ (See figure~\ref{ms_fig4} 
for one of such fitting results). In other words, our $\overline{M_s}$ data are 
consistent with the standard scenario for clean systems. This implies that the 
upper critical dimension of the clean model is not affected by the considered 
configurational disorder. It should be pointed out that the $\overline{M_s}$ 
data obtained here can also be fitted to the ansatz $a_1|j_c-j|^{b_1}$.
Furthermore, the average value of $b_1$ determined from the corresponding
good fits are 0.410(16). Finally the critical points 
$(J'/J)_c$ obtained from the fits of these two ansatzes are given by 
4.166(3) and 4.162(3) on average, respectively. Based on these results,
at this stage we are not able
to reach a definite answers of whether the calculated $\overline{M_s}$
data here receive any logarithmic corrections.  
Later when discussing the determination of $\overline{T_N}$,
we will argue that our data are in favour of the 
scenario that logarithmic corrections do enter the $J'/J$-dependence of the 
related observables. 

\vskip0.5cm

\subsection{The determination of $\overline{T_N}$}

\vskip1cm
\begin{figure}
\vskip0.5cm
\begin{center}
\vbox{
\includegraphics[width=0.38\textwidth]{rhosLJ2.0.eps}\vskip1cm
\includegraphics[width=0.38\textwidth]{Q2J2.0.eps}
}
\end{center}
\caption{$\overline{\rho_s}L$ (top bottom) and $\overline{Q_2}$ (bottom panel) 
as functions of $T/J$ for $J'/J = 2.0$ and
$L$ = 8, 10, 12, 16, 20, 24, 28, 32, 36. $J$ is 1.0 in our calculations. 
The solid lines are added to guide the eye.}
\label{TN_fig1}
\end{figure}


\begin{figure}
\vskip0.5cm
\begin{center}
\vbox{
\includegraphics[width=0.38\textwidth]{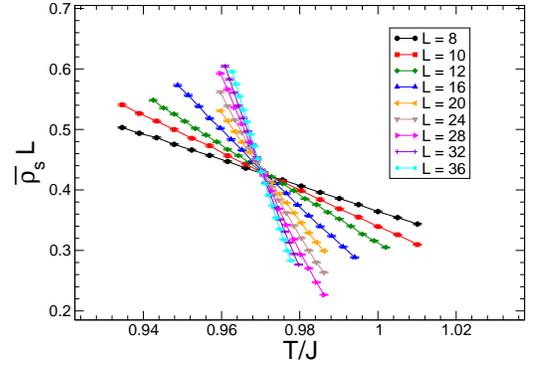}\vskip1cm
\includegraphics[width=0.38\textwidth]{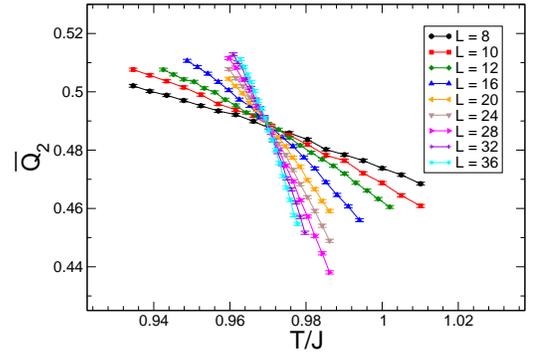}
}
\end{center}
\caption{$\overline{\rho_s}L$ (top bottom) and $\overline{Q_2}$ (bottom panel) 
as functions of $T/J$ for $J'/J = 3.0$ and
$L$ = 8, 10, 12, 16, 20, 24, 28, 32, 36. $J$ is 1.0 in our calculations. 
The solid lines are added to guide the eye.}
\label{TN_fig2}
\end{figure}

\begin{figure}
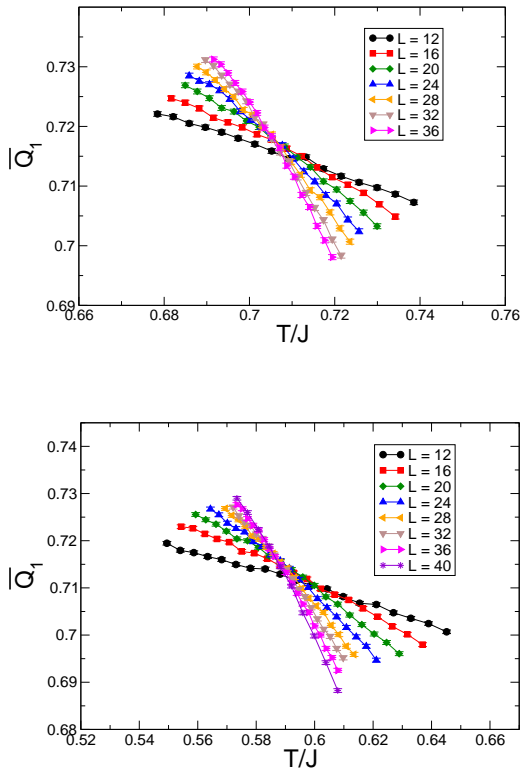

\vskip0.5cm
\begin{center}
\vbox{
\includegraphics[width=0.38\textwidth]{Q1J3.8.eps}\vskip1cm
\includegraphics[width=0.38\textwidth]{Q1J3.95.eps}
}
\end{center}
\caption{$\overline{Q_1}$ as functions of $T/J$ for $J'/J = 3.8$ (top panel) and
$J'/J = 3.95$ (bottom panel). The box sizes $L$ for these two values of
$J'/J$ are $L$ = 8, 10, 12, 16, 20, 24, 28, 32, 36 (and 40 for $J'/J=3.95$). 
$J$ is 1.0 in our calculations. The solid lines are added to guide the eye.}
\label{TN_fig3}
\end{figure}

\begin{figure}
\vskip0.5cm
\begin{center}
\includegraphics[width=0.38\textwidth]{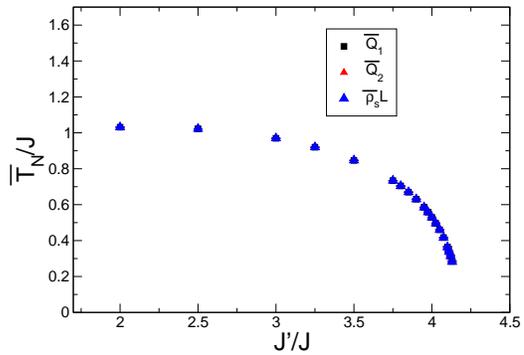}
\end{center}
\caption{$\overline{T_N}$, obtained from $\overline{Q_1}$, $\overline{Q_2}$,
and $\overline{\rho_s}L$, as functions of $J'/J$.}
\label{TN_fig4}
\end{figure}

The employed observables for calculating $\overline{T_N}$ are 
$\overline{\rho_s}L$, $\overline{Q_1}$, as well as $\overline{Q_2}$.
Notice a constraint standard finite-size scaling ansatz of the form 
$(1+b_0L^{-\omega})(b_1 + b_2tL^{1/\nu} +b_3(tL^{1/\nu})^2+...$), up to second,
third, and (or) fourth order in $tL^{1/\nu}$, is adopted to fit the data. 
Here $b_i$ for $i=0,1,2,...$ are some constants and 
$t = \frac{T-\overline{T_N}}{\overline{T_N}}$. For some $J'/J$,  
ansatz up to fifth order in $tL^{1/\nu}$ is used. 
The data of $\overline{\rho_s}L$ and $\overline{Q_2}$ for $J'/J = 2.0$ ($J'/J = 3.0$)  
are shown in fig.~\ref{TN_fig1} (fig.~\ref{TN_fig2}). In addition, the 
$\overline{Q_1}$ data of $J'/J = 3.8$ and $J'/J = 3.95$ are presented in fig.~\ref{TN_fig3}. 
For every $J'/J$, ansatzes of various order in $tL^{1/\nu}$ are employed
to fit several sets of data (Each set of data has different range of $L$).
The cited values of $\overline{T_N}$
in this study are estimated by averaging the corresponding results of good fits.
Furthermore, the error bar of each quoted $\overline{T_N}$ is determined from the uncertainty of every 
individual $\overline{T_N}$ of the associated good fits. 
For this analysis we consider a fit with $\chi^2/{\text{DOF}} \lessapprox 2.0$ a 
good fit. In some cases, more restricted conditions on $\chi^2/{\text{DOF}}$ and the
obtained results are imposed for consistency. 
The determined $\overline{T_N}$ from the observables $\overline{\rho_s} L$,
$\overline{Q_1}$, and $\overline{Q_2}$ are shown in fig.~\ref{TN_fig4} \cite{Tan16}.
In addition to $\overline{T_N}$, other interesting physical quantities to study are
the critical exponents $\nu$ and $\omega$ appearing in the relevant
finite-size scaling ansatzes. Notice the dimensionality as well as some
critical exponents are present in the conventional 
finite-size scaling ansatz involving $\rho_s$. Based on the 
analysis of $\overline{M_s}$ in previous section, while it is plausible
to employ the conventional finite-size scaling ansatz of clean models
for the considered finite-temperature phase transitions, 
one cannot rule out the possibility that when $J'/J$ is close enough to 
the critcal point, the effective dimensions of the system as well as 
the values of the exponents in the scaling ansatz receive corrections 
due to the employed disorder. On the other hand, because of their definition, 
Binder ratios, like $\overline{Q_1}$ and $\overline{Q_2}$ calculated here, 
do not 
encounter such kind of subtlety. Indeed, the values of $\nu$ obtained from the 
fits related to $\overline{\rho_s}L$ are systematic smaller than the 
corresponding 
results associated with $\overline{Q_1}$ and $\overline{Q_2}$. Such a trend is
becoming more clear as one approaches the quantum critical point $(J'/J)_c$. 
Hence here we only 
summarize the results of $\nu$ obtained from fitting the data of 
$\overline{Q_1}$ and $\overline{Q_2}$ to their expected ansatzes 
\cite{Tan16.1}. 
The individual average $\nu$ of $J'/J$ with $J'/J < 4.0$, obtained 
from the related good fits of $\overline{Q_1}$ ($\overline{Q_2}$),
ranging from $0.69$ to $0.72$ (0.69 to 0.73). On the other hand, 
the values of $\nu$ calculated for $4.0 \le J'/J \le 4.12$ ($J'/J = 4.13$) 
lie(s) between 0.63 and 0.69 (0.60 and 0.62). We attribute this result to the fact
that data of large box size are limited for $J'/J \ge 4.0$. 
Such scenario is observed for clean dimerized models as well \cite{Wen08,Jia12}.
Finally, the determination of $\omega$ with reasonable precision
is hindered by the strong 
correlation between $\omega$ and its related prefactor $b_0$ in the fitting
formulas.

\begin{figure}
\begin{center}
\includegraphics[width=0.38\textwidth]{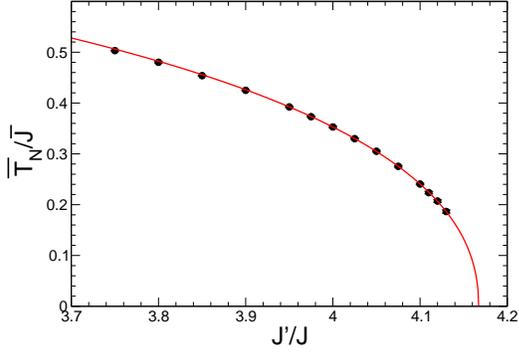}
\end{center}
\caption{Fit of $\overline{T_N}/\overline{J}$ data (obtained from
$\overline{Q_1}$) with 
$J'/J \ge 4.0$ to their theoretical expression with a logarithmic 
correction $a_3|j_c-j|^{b_3}|\ln(|(j_c-j)/j_c|)|^{3/11}$. The $b_3$ obtained from
the fit is given by 0.494(8).  }
\label{TN_fig5}
\end{figure}

After obtaining the numerical values of $\overline{T_N}$, we turn to
the study of whether a logarithmic correction, like the one 
associated with $\overline{M_s}$, exists for $\overline{T_N}$ when 
$\overline{T_N}$
is treated as a function of $J'/J$. Similar to our earlier analysis for $\overline{M_s}$,
we use two ansatzes, namely
\begin{eqnarray}
&&a_2|j_c-j|^{b_2}\nonumber \\
&&a_3|j_c-j|^{b_3}|\ln(|(j_c-j)/j_c|)|^{3/11}
\label{TN_log}
\end{eqnarray}
to fit the data of $\overline{T_N}/\overline{J}$ determined from all
the calculated observables $\overline{Q_1}$, $\overline{Q_2}$ 
and $\overline{\rho}_sL$.
Notice the exponent $b_3$ of the second ansatz of Eq.~(\ref{TN_log})
is predicted to take its mean-field value 0.5.
Furthermore, we investigate the physical quantity 
$\overline{T_N}/\overline{J}$ instead of
$\overline{T_N}$ is motivated by the analysis done in \cite{Yan15}.
The consideration of $\overline{T_N}/\overline{J}$ is also natural since
it is a dimensionless quantity. Interestingly, for all three data sets,
we arrive at good fits ($\chi^2/{\text{DOF}} \le 1.0$) using the second ansatz 
of Eq.~(\ref{TN_log}) when data points of $\overline{T_N}/\overline{J}$ with 
$J'/J \ge 3.75$ are included in the fits. 
On the other hand, the results obtained  
from applying the first ansatz to fit the data
have much worse fitting quality. As fewer data are included
in the fits, while the results related to the second ansatz remain good, 
the $\chi^2/{\text{DOF}}$ associated with the fits employing the first ansatz 
continue to be very large (except those of the fits using data 
sets close to $(J'/J)_c$). 
Notice ocassionally fits with the first ansatz 
lead to good results, but not in a systematic manner.  
The exponent $b_3$ and the critical point $(J'/J)_c$
obtained from all the good fits are given by 0.49(1) and 4.166(2) 
on the average, respectively.
The calculated value of $b_3$, namely $b_3 = 0.49(1)$ is in reasonably good 
agreement with the expected mean-field result 0.5. 
A fit of this analysis including the logarithmic correction
is shown in fig.~\ref{TN_fig5}.  
According to what have been reached so far, we conclude that our data are fully 
compatible 
with the scenario that the upper critical dimension, associated with the 
relevant quantum phase transition of our model, is the same as that of the 
corresponding clean model. In particular, the related critical exponents are 
in agreement with the theoretical predictions of clean systems.


\begin{figure}
\vskip1cm
\begin{center}
\vbox{
\includegraphics[width=0.38\textwidth]{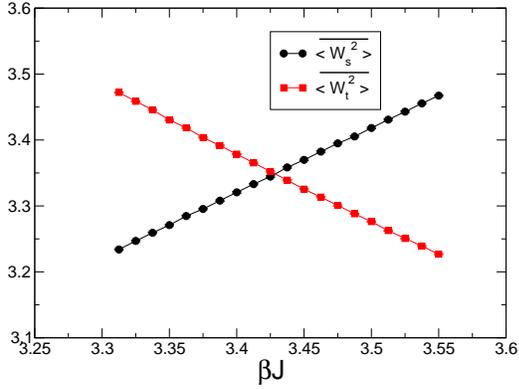}\vskip1cm
\includegraphics[width=0.38\textwidth]{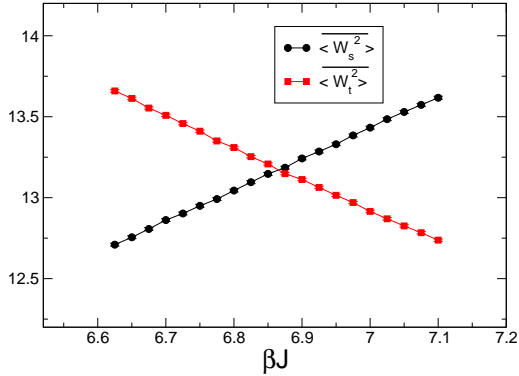}
}
\end{center}
\caption{The spatial and temporal winding numbers squared, as functions of
$\beta J$ at $J'/J = 3.0$, for $L=8$ (top panel) and $L=16$ (bottom panel). 
$J$ is 1.0 in our simulations.}
\label{c_fig1}
\end{figure}


\subsection{The determination of $\overline{c}$}

\begin{figure}
\vskip0.5cm
\begin{center}
\vbox{
\includegraphics[width=0.38\textwidth]{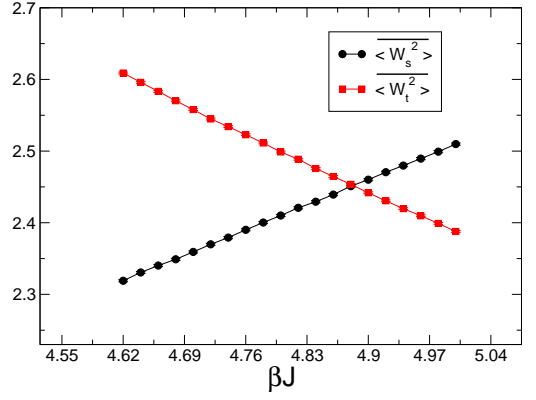}\vskip1cm
\includegraphics[width=0.38\textwidth]{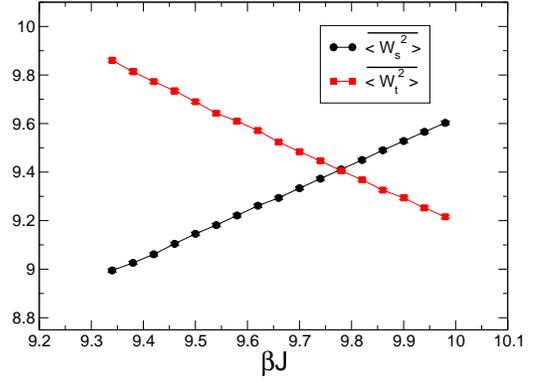}
}
\end{center}
\caption{The spatial and temporal winding numbers squared, as functions of
$\beta J$ at $J'/J = 3.9$, for $L=12$ (top panel) and $L=24$ (bottom panel). 
$J$ is 1.0 in our simulations.}
\label{c_fig2}
\end{figure}

\begin{figure}
\vskip0.5cm
\begin{center}
\includegraphics[width=0.38\textwidth]{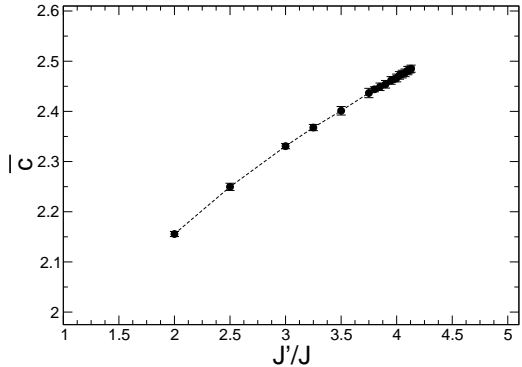}
\end{center}
\caption{The estimated values of $\overline{c}$ for all the considered $J'/J$.}
\label{c_fig3}
\end{figure}

The values of spinwave velocity $\overline{c}$ are
estimated using the idea of winding numbers squared as suggested
in Refs.~\cite{Jia11,Sen15}. Specifically, for a given $J'/J$ and a box size $L$,
one varies the inverse temperature $\beta$ so that the spatial 
and temporal winding numbers squared 
($\overline{\langle W_s^2 \rangle} = \frac{1}{3}(\overline{\langle W_1^2 \rangle + \langle W_2^2 \rangle + \langle W_3^2 \rangle}$) 
and $\overline{\langle W_t^2\rangle }$) take the same values. Assuming
one reaches the condition $\overline{\langle W_s^2 \rangle} = \overline{\langle W_t^2\rangle} $
at an inverse temperature $\beta^{\star}$, then the spinwave velocity 
$\overline{c}(J'/J,L)$ corresponding to this set of parameters
$J'/J$ and $L$ is given by $\overline{c}(J'/J,L) = L/\beta^{\star}$.
Notice with our implementaion of configurational disorder,
the three spatial winding numbers squared take the same 
values after one carries out the disorder average.
For the $J'/J$ of smaller magnitude, the convergence of $\overline{c}$
to their infinite volume values are checked using the data of $L=8$ and 
$L=16$. In addition, the bulk $\overline{c}$ for large 
magnitude $J'/J$ are obtained from the data of $L=12$ and $L=24$. With the
statistics reached here, we find
that for all the considered $J'/J$ the corresponding bulk spinwave velocities 
$\overline{c}$ can be correctly given by the results at $L=16$ or $L=24$. 
The convergence of the spinwave wave velocities to their bulk values
for $J'/J = 3.0$ and $J'/J = 3.9$ are demonstrated in figs. \ref{c_fig1} 
and \ref{c_fig2}, 
respectively, and the bulk $\overline{c}$ we obtain are presented in 
fig. \ref{c_fig3}. We would like to point out that for 
$J'/J \ge 4.0$, our estimated central values of $\overline{c}$ for $L=12$ and 
$L=24$ differ by only less than 0.34 percent and are within their corresponding 
error bars. Therefore the results of $\overline{c}$ shown in fig. \ref{c_fig3} 
should be very reliable.

\vskip2cm

\subsection{The scaling relations $\overline{T_N}/\overline{c}^{3/2} = A \overline{M_s}$
and $\overline{T_N}/\overline{J} = A_1 \overline{M_s}$}

\begin{figure}
\vskip1.5cm
\begin{center}
\vbox{
\includegraphics[width=0.38\textwidth]{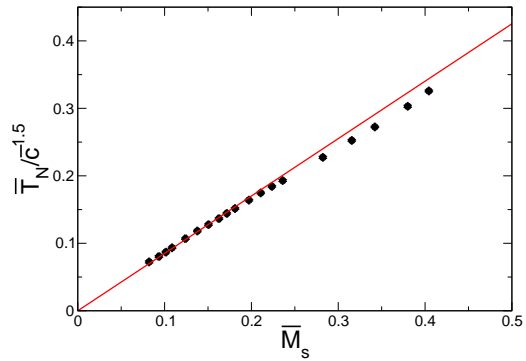}\vskip0.8cm
\includegraphics[width=0.38\textwidth]{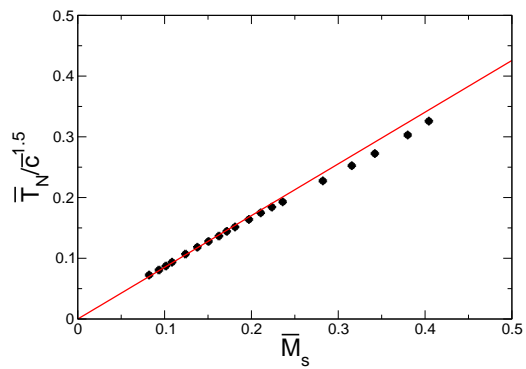}
}
\end{center}
\caption{Fits of $\overline{T_N}/\overline{c}^{3/2}$ as linear functions of $\overline{M_s}$
passing through the origin. The $\overline{T_N}$ of the top and bottom panels are
obtained from $\overline{Q_1}$ and 
$\overline{Q_2}$, respectively.}
\label{TNcM_fig1}
\end{figure}

\begin{figure}
\vskip1cm
\begin{center}
\includegraphics[width=0.38\textwidth]{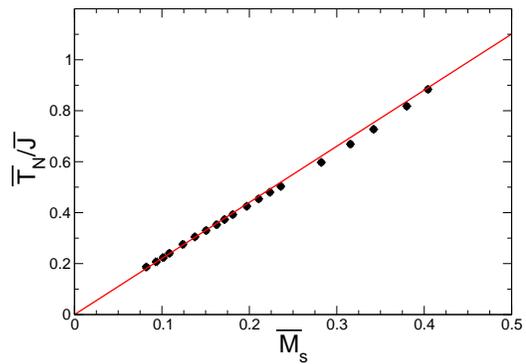}
\end{center}
\caption{Fit of $\overline{T_N}/\overline{J}$ as a linear function of $\overline{M_s}$
passing through the origin. The used values of $\overline{T_N}$ are
obtained from $\overline{Q_1}$.
The average of the slopes for all the good fits including those resulting from
$\overline{Q_2}$ and $\overline{\rho_s}L$ is given by 2.22(1).}
\label{TNbarJM_fig2}
\end{figure}

Having obtained $\overline{M_s}$, $\overline{T_N}$, and $\overline{c}$, we move
to examine whether the scaling relations $\overline{T_N}/\overline{c}^{3/2} = A \overline{M_s}$
and $\overline{T_N}/\overline{J} = A_1 \overline{M_s}$,
which are confirmed for clean system(s), remain true for the model with
the introduced configurational disorder studied here. Actually, the validity 
of these relations
for our model is expected, since our data of $\overline{T_N}/\overline{J}$ and $\overline{M_s}$
are fullly compatible with the theoretical predictions for clean systems, and neither
$\overline{c}$ nor $\overline{J}$ receives any logarithmic corrections. 
Indeed, the obtained results of $\overline{T_N}/\overline{c}^{3/2}$, when 
being treated as a
function of $\overline{M_s}$, can be fitted to the ansatz of
$A\overline{M_s}$ using the data with the corresponding $\overline{M_s}$ having small magnitude. 
Furthermore, the prefactor $A$ determined from the fits has an 
average $A_{\text{avg}}\sim 0.858(4)$. 
Two outcomes of the fits are demonstrated in fig. 
\ref{TNcM_fig1}. Notice both the uncertainties of $\overline{T_N}/\overline{c}$
and $\overline{M_s}$ are taken into account in the fits. 
Similarly, close to the quantum phase transition, our data of
$\overline{T_N}/\overline{J}$ and $\overline{M_s}$ do satisfy a linear
relation as well, see fig.~\ref{TNbarJM_fig2}.

One may wonder whether the data of $\overline{T_N}/\overline{c}^{3/2}$ can be fitted to the ansatz 
of the form $A\overline{M_s} + B$ with $B$ being consistent with zero. 
We have applied such analyses to the data obtained close to $(J'/J)_c$.
In particular, we arrive at the result that the constants $B$ determined from 
the fits satisfy $|B| \lesssim 0.005$ and the magnitude 
of corresponding uncertainties are comparable with $|B|$. We consider these 
outcomes 
as a strong indication that our data of $\overline{T_N}/\overline{c}^{3/2}$, as a function 
of $\overline{M_s}$, can be described well by a linear function of $\overline{M_s}$ passing 
through the origin.

\section{Discussions and Conclusions}

\begin{figure}
\vskip1cm
\begin{center}
\includegraphics[width=0.38\textwidth]{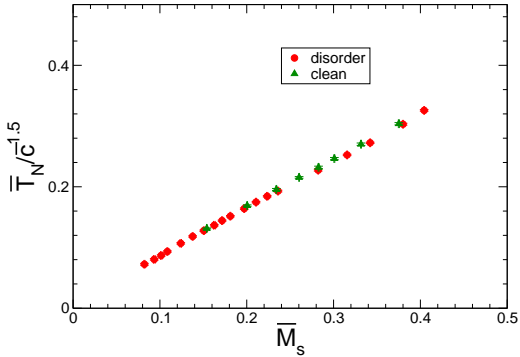}
\end{center}
\caption{$\overline{T_N}/\overline{c}^{3/2}$ as functions of $\overline{M}$ for both the considered disordered 
model in this investigation and the clean model studied in Ref.~\cite{Kao13}.
The $\overline{T_N}$ used in the figure for the disordered model 
are determined from the observable $\overline{Q_1}$.}
\label{conclusion_fig1}
\end{figure}

\begin{figure}
\vskip1cm
\begin{center}
\includegraphics[width=0.38\textwidth]{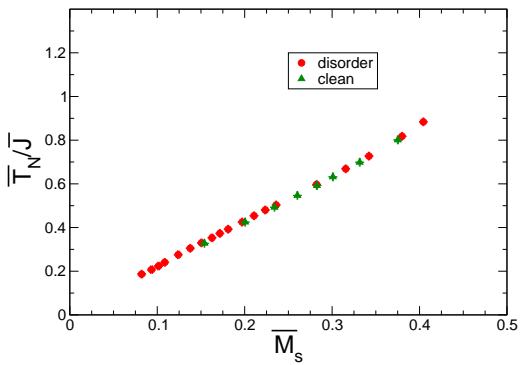}
\end{center}
\caption{$\overline{T_N}/\overline{J}$ as functions of $\overline{M}$ for both the considered disordered 
model in this investigation and the clean model studied in Ref.~\cite{Kao13}.
The $\overline{T_N}$ used in the figure for the disordered model 
are determined from the observable $\overline{Q_1}$.}
\label{conclusion_fig2}
\end{figure}

For clean 3D dimerized quantum Heisenberg models, it is established that 
the physical quantities $T_N/c^{3/2}$ and $T_N/\overline{J}$, as functions of
$M_s$, scale linearly with $M_s$. Notice since three spatial dimensions is 
the upper critical dimensions associated with the related quantum phase 
transition, one expects there are logarithmic corrections to $T_N$ and $M_s$ 
close to the critical point. The linear scaling relations
between $T_N$ and $M_s$ indicate that the exponents associated with the 
logarithmic corrections
to $T_N$ and $M_s$ take the same values. This result is obtained theoretically
in Refs. \cite{Ken04,Ken12} and is confirmed in Ref. \cite{Yan15}. 

Motivated by these universal scaling relations 
between $T_N$ and $M_s$ for the clean
3D dimerized systems, here we study these relations for a 3D quantum Heisenberg 
model with configurational disorder. A remarkable result observed in our 
investigation is that close to the considered critical point,
the relations $T_N/c^{3/2} = A M_s$ and 
$T_N\overline{J} = A_1M_s$ remain valid for
the studied model with the introduced disorder. 
In addition, both the obtained data of $\overline{T_N}$ and $\overline{M_s}$
in this study do receive multiplicative logarithmic corrections and  
are fully compatible with the expected critical behaviour for 
clean dimerized models. This indicates that
the related quantum phase tranistion of the studied system may be described by
the same critical theory as that of its clean analogues. 
In Ref.~\cite{Yan15},
it is found that for the clean 3D double cubic quantum Heisenberg model the 
numerical value of the coefficient $A$ in 
$T_N/c^{3/2} = A M_s$ is given by  $A \sim 1.084$. For the disordered model 
considered here, the corresponding coefficient is given by $A \sim 0.86$, 
which is different from $A \sim 1.084$ associated with the double cubic model 
studied in Ref.~\cite{Yan15}. 
Consequently, this coefficient $A$ is likely not universal and depends on 
the microscopic details of the investigated systems. Interestingly, using the 
data determined here and that of a clean system calculated in 
Ref.~\cite{Kao13}, smooth universal curves  
associated with the studied scaling relations do show up
(See figs.~\ref{conclusion_fig1} and \ref{conclusion_fig2}). 
This implies that while not being
generally true, universal coefficients may still exist within models sharing
some similar characters. It will be interesting to investigate, in a more 
quantitative manner, that under what conditions will the coefficients $A$ and
$A_1$ of different models take the same numerical values. Finally, although 
based on our investigation we conclude that the coefficient $A$ is likely not 
universal, we find the quantity $A/(J'/J)_c^{3/2}$ determined in this 
investigation matching very well with the corresponding result 
in \cite{Yan15}. Specifically, using the data available in Ref.~\cite{Yan15} 
and here, we reach 
$A/(J'/J)_c^{3/2} \sim 0.1019$ and $A/(J'/J)_c^{3/2} \sim 0.1011$ for the double 
cubic model and the disordered model studied here, respectively. Notice the 
numerical values of $A/(J'/J)_c^{3/2}$ obtained from two different models are 
in very good agreement with each other. While our 
preliminary result of $A/(J'/J)_c^{3/2}$ of other dimerized model does not seem  
to support the scenario that the quantity $A/(J'/J)_c^{3/2}$ takes
a universal value, the investigation carried out in this study suggests 
true universal quantities may still emerge for both the clean and disordered 
3D antiferromagnets (Which share some similar properties). To uncover the 
possible hidden universal relations
of 3D antiferromagnets will be an interesting topic to conduct in the future.


\section{Acknowledgments}
\vskip-0.3cm
We thank A.~W.~Sandvik for encouragement. This study is partially supported by 
MOST of Taiwan.


\begin{thebibliography}{99}
\bibitem{Tan16}
In an early version of this manuscript, we claimed that for
the $J'/J$ close to $(J'/J)_c$, the corresponding values of 
$\overline{T_N}$ determined from $\overline{\rho_s} L$
deviate slightly from those associated with the Binder ratios. 
With higher order ansatzes, careful investigation of deciding 
data range for the fits, and more statistics, 
now the discrepancies disappear for most of the considered 
$J'/J$. Consequently, some numerical analysis are redone and the
corresponding results are updated. Conclusions remain the same.
\bibitem{Tan16.1}
While it is likely that the determination of $\nu$ by considering
the observable $\overline{\rho_s}L$ is affected
by the presence of bond disorder,
the quantitative agreement among the $\overline{T_N}$
obtained from $\overline{\rho_s}L$, $\overline{Q_1}$, and $\overline{Q_2}$
indicates that the calculated results of $\overline{T_N}$ 
are reliable.
\end{thebibliography}
\end{document}